\documentclass[pra,preprint,amsmath,amssymb,showpacs,showkeys]{revtex4}
\usepackage{bm}
\usepackage{graphics}

\newcommand{\vare}{\varepsilon }
\newcommand{\md}{\mathrm{d}}
\newcommand{\eqb}{\begin{equation}}
\newcommand{\eqe}{\end{equation}}
\newcommand{\pd}{\partial}

\begin{document}

\title{Asymmetries in the tunneling probability of Bose-Einstein condensate 
in an accelerating optical lattice}
\author{Valery S. Shchesnovich }
\author{Solange B. Cavalcanti}
\affiliation{ Departamento de F\'{\i}sica - Universidade Federal
de Alagoas, Macei\'o AL 57072-970, Brazil }

\begin{abstract}
We derive a two-band finite-dimensional model for description of the condensate
tunneling in an accelerating optical lattice, taking into account the fine Bloch
band structure. The  model reveals a very strong dependence of the final band
populations on the initial populations and phases. Most importantly, additionally
to the known asymmetric dependence on the nonlinearity, there is also a notable
asymmetry in the sensitivity of the tunneling probability to the
nonliearity-induced initial population of the Bloch band to which the tunneling
takes place. This fact can explain the experimentally observed  unexpected
independence of the upper-to-lower tunneling probablity on the nonlinearity.
Finally, we compare the predictions of the two-band model with that of the
well-known nonlinear Landau-Zener model and find disagreement  when the two bands
are initially populated. The disagreement can be qualitative and reveals itself
even for a negligible nonlinearity. However, the two models agree remarkably well
if just one band is populated initially.

\end{abstract}

 \maketitle

\section{Introduction}
\label{SECI}
Since the successful realization of Bose-Einstein condensation
(BEC) in an optical lattice \cite{BEC1,BEC2} the experimental
efforts are directed to explore and understand the behavior of BEC
in a periodic potential. Many of the BEC phenomena in an optical
lattice have an analog  in the solid state physics. For instance,
Bloch oscillations  and Landau-Zener tunneling of BEC were
experimentally observed \cite{BLLZ1,BLLZ2,LZ3,LZ4} in an
accelerating lattice. The atomic interactions in BEC, however,
give rise to the essentially nonlinear effects such as the
asymmetry of the tunneling probability between the two adjacent
bands \cite{LZ3,LZ4} and the Landau and dynamical instabilities
\cite{LZ4,IN1,IN2,IN3,IN4}. Accordingly, two essentially different
regimes of BEC  dynamics  in an optical lattice can be identified
\cite{KKSMOD,LZ4}: the ``instability regime'', which realizes for
a small to intermediate values of the lattice acceleration and the
``Landau-Zener regime'', which realizes for larger accelerations.
When the acceleration is small, the condensate slowly crosses the
edge of the Brillouin zone allowing the unstable modes to dominate
the order parameter \cite{NJP,KKSMOD}. For instance, in one of the
bands the modulational instability \cite{MODIN} will result in the
bright soliton formation  even in a repulsive BEC \cite{BRSOL}.

For the ``Landau-Zener regime'' a reduced two-level model (the
nonlinear Landau-Zener model) was proposed in Refs. \cite{WN,ZG}
to account for the effect of the atomic interactions on the
tunneling probability of BEC (see also Refs. \cite{NLLZ,NJP} for
further development). This model is  derived for a weak optical
lattice  by an analog of the method of free electrons in the solid
state theory (see for instance, Ref. \cite{Ziman}) and is
essentially the Landau-Zener model \cite{LZMOD} modified by the
nonlinear terms due to the atomic interactions in BEC.

The predictions of the nonlinear Landau-Zener model were
experimentally tested \cite{LZ3,LZ4}. In accordance with the
theoretical prediction the atomic interactions in BEC make the
tunneling probability asymmetric, but still a significant
disagreement was discovered. Hence, a more accurate model is in
order for description of the condensate tunneling.

To account for the modulational instability present in one of the Bloch bands a
spatio-temporal model was proposed recently \cite{KKSMOD}. Essentially, it consists
of two linearly coupled nonlinear Schr\"odinger equations for the slowly varying
amplitudes of the two Bloch waves bordering the Brillouin zone edge (i.e.  at
$k=k_B$). The key role of the modulational instability of Bloch wave amplitudes in
the transition from the reversible tunneling to the asymmetry in the band
populations is revealed. Therefore, this model is suitable for description of the
``instability regime'', when the Brillouin zone edge crossing time is comparable to
the characteristic time of development of the modulational instability. However,
its applicability is restricted to the case of a very small lattice acceleration
and thus it cannot be applied to the ``Landau-Zener regime'' (as defined above).

The aim of this paper is to propose an improved finite-dimensional system (the
two-band model) for  BEC tunneling in an accelerating optical lattice. We take into
account  the fine structure of the Bloch bands, which is of crucial importance if
the initial conditions are arbitrary. Only the ``Landau-Zener regime'' is
considered. Hence, we can safely neglect the spatial inhomogeneity of the Bloch
wave amplitudes. We expand the BEC order parameter over the Bloch waves with a
time-dependent wave index. Thus our method is similar to Houston's approach
\cite{HOUST} to the accelerating electrons in the solid state theory. The resulting
finite-dimensional model can also have application in the study of ultracold atoms
in a periodic potential submitted to a constant external force \cite{CA1,CA2}.

There are two main results. First, the two-band model reveals a dramatic dependence
of the final Bloch band populations on the initial populations and the phase
difference. Most importantly, additionally to the asymmetry in the dependence  of
the tunneling probability on the nonlinearity strength, we reveal also the
asymmetry in the sensitivity of  the tunneling probability  to the population of
the adjacent Bloch band (i.e. to which the tunneling takes place). This fact,
combined with the nonlinearity-induced population of the adjacent band, can explain
the independence of the upper-to-lower tunneling probability on the nonlinearity
suggested by the experimental results of Ref. \cite{LZ3}. Second, there is a
significant disagreement between the predictions of the two-band model and the
Landau-Zener model if the two bands are initially populated. The disagreement can
be qualitative and reveals itself even for a negligible nonlinearity. The two
models agree in the predictions if only one Bloch band is populated initially.

The paper is organized as follows. In the next section a derivation of the two-band
model is presented. Some details of the derivation are relegated to appendix
\ref{APPA}. Then, in section \ref{SECIII}, the case of a weak lattice is
considered. The predictions of the  two-band model are compared to those of the
nonlinear Landau-Zener model in section \ref{SECIV}. The
 two-band model in the diabatic basis is given in
appendix \ref{APPB}. Section \ref{SECV} contains discussion of the
asymmetric BEC tunneling. Finally,  a general discussion and
perspectives are contained in the concluding section \ref{SECVI}.


\section{Derivation of the two-band model}
\label{SECII}
We consider a quasi one-dimensional BEC, i.e. the transverse width is much smaller
than the length of the condensate.  For not too large number of atoms the
transverse dynamics of BEC is that of a quantum particle confined to the lower
energy level of the parabolic trap \cite{BOOK}. In terms of the $s$-wave scattering
length $a_s$, the transverse oscillator length $\ell_\perp =
\sqrt{\hbar/m\omega_\perp}$ and the linear condensate density $n_{1D} = N_d/d$,
where $N_d$ is the number of atoms per lattice site, the condensate can be
considered as quasi one-dimensional under the  condition $n_{1D} a_s \ll 1$
\cite{KinTerm}. In the experiments  the usual values are:  $d\sim 1\mu$m and
$a_s\sim 1$nm. Hence, to satisfy the quasi one-dimensionality condition one has to
have $N_d \ll 1000$. The BEC order parameter $\Phi(\vec{r}_\perp,x,t)$ then can be
approximated as a product of the wave-function for the ground state of the
transverse parabolic trap and a function describing the longitudinal motion of the
condensate:
\eqb \Phi(\vec{r}_\perp,x,t) =
\frac{1}{\sqrt{\pi}\ell_\perp}\exp\left\{-\frac{r_\perp^2}{2\ell_\perp^2}\right\}
\psi(x,t).
\label{EQ1}\eqe
For the reduced order parameter $\psi(x,t)$ the Gross-Pitaevskii
equation describing the condensate in an accelerated optical
lattice  reads
\eqb
i\hbar\pd_t\psi = -\frac{\hbar^2}{2m}\pd^2_x \psi +\left(
V_0\cos(2k_L [x-x_0(t)]) +
\frac{m\omega^2_\parallel}{2}x^2\right)\psi +
\frac{gn_{1D}}{2\pi\ell^2_\perp} |\psi|^2\psi,
\label{EQ2}\eqe
where $g = 4\pi\hbar^2a_s/m$, $k_L$ is the optical lattice index
and $x_0(t)$ describes the acceleration of the lattice.  Here the
order parameter is normalized as follows $d^{-1}\int_0^d \md x\,
|\psi|^2 = 1$ with $d = \pi/k_L$ being the lattice spacing. The
following dimensionless variables will be used below:
\eqb
\tilde{t} = \frac{8E_R}{\hbar}t,\quad \tilde{x} =
2k_L[x-x_0(t)],\quad {v}_0 = \frac{V_0}{8E_R},\quad c = \frac{ a_s
n_{1D}}{2\ell^2_\perp k_L^2}, \quad \lambda^2 = (2\ell_\parallel
k_L)^{-2},
\label{EQ3}\eqe
where  $E_R = {\hbar^2k^2_L}/(2m)$ is the recoil energy and
$\ell_\parallel = \sqrt{\hbar/m\omega_\parallel}$.   We have
\eqb
i\pd_{\tilde{t}}\tilde{\psi} = \frac{1}{2}\left(-i\pd_{\tilde{x}}
- \frac{\md \tilde{x}_0}{\md\tilde{t}}\right)^2\tilde{\psi}
+\left(v_0\cos(\tilde{x})+\frac{\lambda^2}{2}[\tilde{x}+\tilde{x}_0]^2
\right)\tilde{\psi}+ c|\tilde{\psi}|^2\tilde{\psi},
\label{EQ4}\eqe
where $\tilde{x}_0 = 2k_Lx_0$ and
\[
\tilde{\psi} = \exp\left\{-\frac{i}{2}\int\limits^{\tilde{t}}_0\md
\tau\left(\frac{\md \tilde{x}_0}{\md
\tilde{t}}\right)^2\right\}\psi.
\]
The normalization now reads $(2\pi)^{-1}\int_0^{2\pi}\md
\tilde{x}\,|\tilde{\psi}|^2 = 1$.  As we will use only the
dimensionless variables of equation (\ref{EQ4}), we drop all the
tildes from $x$, $t$, and $\psi$, for simplicity.

We assume that the quasi-momentum spread  of the condensate in the
optical lattice is negligible compared to the width of the
Brillouin zone. One necessary condition is that the optical
lattice be long enough, i.e. the parabolic trap contains many
lattice periods: $\lambda\ll 1$. Then, the effect of the parabolic
trap is negligible. We set $\lambda=0$.

An additional condition is that the time of the Brillouin zone
edge crossing  is much less that the characteristic time of the
modulational instability development. This sets a lower limit on
the acceleration $\alpha$. The latter time can be estimated as
follows. In the slowly varying envelope approximation
\cite{MODIN,KKSMOD} the spatial Bloch amplitude will be governed
by a nonlinear Schr\"odinger equation with  an effective
interaction coefficient $c\chi_{nn}$. The characteristic time of
the instability development is the inverse of the maximal growth
rate $t_{MI}\sim |c|^{-1}$.

Formation of gap solitons is also possible for a nonlinearity of
the order of the gap between the Bloch bands \cite{GAP}. In the
following, however, we consider only  the weak nonlinearity limit
$c\ll v_0$, used in the experiment \cite{LZ3}.

Under the conditions that the condensate has a narrow initial
momentum spread and the time of the Brillouin zone edge crossing
is much less that the characteristic time of the modulational
instability development the solution to equation (\ref{EQ4}) can
be expanded in the basis of the Bloch waves with a definite
(time-dependent) index $q(t)$ and an overall linear phase due to
the acceleration:
\eqb
\psi = e^{i\mathrm{v}(t)x}\sum_{n=1}^\infty
A_n(t)\Psi_{n,q(t)}(x),
\label{EQ5}\eqe
where $\mathrm{v}(t)\equiv \dot{ x}_0$ and $\Psi_{n,q}(x)\equiv
e^{iqx}u_{n,q}(x)$ satisfies
\eqb
\left\{-\frac{1}{2}\pd^2_x + v_0\cos(x)\right\}\Psi_{n,q}(x) =
E_n(q)\Psi_{n,q}(x).
\label{EQ6}\eqe
Due to the normalization and orthogonality of the Bloch functions,
$(2\pi)^{-1}\int\limits_{-\pi}^\pi\md
x\,\Psi^*_{n,k}(x)\Psi_{m,k}(x)= \delta_{nm}$, the amplitudes
satisfy the condition $\sum\limits_{n=1}^\infty |A_n|^2 = 1$.

It is shown below that, in a vicinity of the Brillouin zone edge
$q = k_B$, for a small acceleration and  under the condition
$E_3(k_B)-E_2(k_B)\gg E_2(k_B)-E_1(k_B)$  one can neglect the
effect of the upper bands on the tunneling between the first two
Bloch bands. The nonlinear term with $c\ll v_0$ can be taken into
account by using a perturbation theory.

The expansion (\ref{EQ5}) is in the spirit of Houston's approach
in the theory of electrons accelerating in a crystal lattice
\cite{HOUST}. It takes into account, for instance, that in the
experiments \cite{LZ3,LZ4} the condensate was in a Bloch state
before the acceleration was turned on.

\subsection{The linear limit $c=0$}
\label{SECII1}
For the moment let us drop the nonlinear term and consider the
linear limit of equation (\ref{EQ4}). The contribution from  the
nonlinear term is accounted later on. Substituting the expression
(\ref{EQ5}) into equation (\ref{EQ4}) and dropping the overall
imaginary exponent gives
\[
-\left[\frac{\md q}{\md t} +\frac{\md \mathrm{v}}{\md
t}\right]x\sum_n A_nu_{n,q}(x) + \sum_ni\frac{\md A_n}{\md
t}u_{n,q}(x) + i\frac{\md q}{\md t}\sum_n A_n \pd_q u_{n,q}(x)
\]
\eqb
= \sum_n E_n(q) A_n u_{n,q}(x).
\label{EQ7}\eqe
Setting $q(t) = q_0 -\mathrm{v}(t)$ to cancel the non-periodic in
$x$ term  and projecting the resulting equation onto the basis
function $u_{n,q}(x)$ we get an equation for the amplitude
\eqb
i\frac{\md A_n}{\md t} = E_n(q)A_n + i\alpha(t)\sum_m A_m \langle
u_{n,q}|\pd_q u_{m,q}\rangle,
\label{EQ8}\eqe
where $\alpha = \ddot{x}_0$. Here the inner product over the
lattice period is used: $\langle f_1|f_2\rangle \equiv
(2\pi)^{-1}\int^{\pi}_{-\pi}\md x\, f_1^*(x)f_2(x)$ with $\langle
u_{n,q}|u_{m,q}\rangle = \delta_{n,m}$.

In appendix \ref{APPA} it is shown that for any periodic potential
$v = v(x)$, with $v(x)$ being an even function,  $u^*_{n,q}(x) =
u_{n,q}(-x)$ and, hence, the diagonal inner product $\langle
u_{n,q}|\pd_q u_{n,q}\rangle =0$, while the non-diagonal is given
as
\eqb
\kappa_{nm}(q)\equiv \langle u_{n,q}|\pd_q u_{m,q}\rangle =
-i\frac{\langle u_{n,q}|\pd_x u_{m,q}\rangle}{E_m(q) - E_n(q)} =
-i\frac{\langle u_{n,q}|\frac{\md v(x)}{\md x}|
u_{m,q}\rangle}{[E_m(q) - E_n(q)]^2},\quad n\ne m.
\label{EQ9}\eqe
The coupling coefficient $\kappa_{nm}$ is real  and satisfies
$\kappa_{nm}(q) = -\kappa_{mn}(q)$.

Note that the coupling coefficient  between the Bloch bands is
inversely proportional to the band level spacing $E_m(k)-E_n(k)$,
since $\kappa_{nm}(k) =\mathcal{O}\left( v_0/[E_m(k) -
E_n(k)]^2\right)$. Thus, for a small acceleration and an optical
lattice satisfying  $E_3(k_B)-E_2(k_B) \gg E_2(k_B) - E_1(k_B)$,
the influence of the upper bands on BEC tunneling between the
first two Bloch bands at  the Brillouin zone edge is negligible.
For instance, the above condition is satisfied by a weak optical
lattice (see section \ref{SECIII}).

Dropping the terms corresponding to the upper bands from equation
(\ref{EQ8}) and making the transformation $A_1 =
\exp\left\{-i\int^t_0\md \tau\,\bar{E}(q(\tau))\right\} a_1$ and
$A_2 = \exp\left\{-i\int^t_0\md \tau\,\bar{E}(q(\tau))\right\}
(-i)a_2$, with $\bar{E}=(E_1 + E_2)/2$, we obtain the two-band
model for the linear case:
\eqb
i\frac{\md a_1}{\md t} =-\vare(q)a_1+\alpha\kappa_{12}(q)a_2,
\label{EQ10}\eqe
\eqb
i\frac{\md a_2}{\md t} = \vare(q)a_2+\alpha\kappa_{12}(q)a_1.
\label{EQ11}\eqe
Here we have used that $\kappa_{21}(q) = -\kappa_{12}(q)$ and set $\vare(q)=
(E_2(q) - E_1(q))/2$. Recall that $\alpha(t) = \ddot{x}_0$ is the lattice
acceleration and $q(t) =q_0-\dot{x}_0$ is a time-dependent parameter. The system of
equations (\ref{EQ10})-(\ref{EQ11}) is consistent with the two-band approximation,
since it conserves the total number of atoms in the two bands: $|a_1(t)|^2
+|a_2(t)|^2 = 1$.

The linear two-band model (\ref{EQ10})-(\ref{EQ11}) is important
in its own right, since it can have direct application in the
study of ultracold atoms in a periodic potential submitted to a
constant external force \cite{CA1,CA2}.

\subsection{Weak nonlinearity: $c\ll v_0$ }
\label{SECII2}
In the experiments on BEC tunneling in an optical lattice
\cite{LZ3,LZ4} the nonlinearity satisfies the condition $c\ll
v_0$. In this case, the nonlinear term can be treated as a
perturbation of the optical lattice potential. Here it is
pertinent to mention that the nonlinear term was also treated by
introduction of an effective optical lattice potential
\cite{EFFPOT,LZ4}. This treatment is of limited validity.

Under the condition $c\ll v_0$ the nonlinear term preserves the
qualitative Bloch band structure, but modifies the Bloch functions
and the band energies. The nonlinear term serves as an additional
lattice potential (in general, time-dependent), the same for all
eigenfunctions. In fact, if we define the nonlinear modes
$\hat{u}_{n,k}(x,t)$ as follows
\eqb
 \left\{\frac{1}{2}(-i\pd_x - k)^2
+ v_0\cos(x)+ c\left|\sum_{m=1}^\infty A_m(t)
\hat{u}_{m,k}(x,t)\right|^2\right\}\hat{u}_{n,k}(x,t) =
\hat{E}_n(k,t)\hat{u}_{n,k}(x,t),
\label{EQ12}\eqe
then the effect of the nonlinear term    can be treated by using
the usual perturbation theory for the eigenvalues $\hat{E}_n(k,t)$
of a perturbed ($(k,t)$-dependent) linear operator. Notice that
the nonlinear modes $\hat{u}_{n,q}$ are orthogonal, $\langle
\hat{u}_{m,k} |\hat{u}_{n,k}\rangle = \delta_{n,m}$, and for $c=0$
coincide with the respective Bloch functions. Thus, for small $c$,
they constitute a complete orthogonal basis (in this respect we
generalize the approach of Refs. \cite{Serk,Kivsh}). In the first
order of the perturbation theory we get
\[
\hat{E}_n(k,t) = E_n(k) + c \langle
u^*_{n,q}(x)\left|\sum_{m=1}^\infty A_m(t)
u_{m,q}(x)\right|^2u_{n,q}(x)\rangle + \mathcal{O}(c^2),
\]
with $\langle ...\rangle = (2\pi)^{-1}\int^{\pi}_{-\pi}\md x
(...)$, while the eigenfunctions $\hat{u}_{n,q}(x) = u_{n,q}(x) +
\mathcal{O}(c)$ which is a sufficient  approximation for the
following. In the two-band approximation we get the following
expressions for the energies of the first two Bloch bands (the
variable $t$ is omitted for simplicity):
\[
\hat{E}_1(k)= E_1(k) + c\left\{\chi_{11}(k)|A_1|^2
+\chi_{12}(k)|A_2|^2 + \sigma_{12}(k)A_1A_2^* +
\sigma^*_{12}(k)A^*_1A_2\right\},
\]
\eqb
\hat{E}_2(k)= E_2(k) + c\left\{\chi_{22}(k)|A_2|^2
+\chi_{12}(k)|A_1|^2 + \sigma_{21}(k)A_2A_1^* +
\sigma^*_{21}(k)A^*_2A_1\right\}.
\label{EQ13}\eqe
Here
\[
\chi_{nn} = \langle|u_{n,k}|^4\rangle,\quad
\chi_{12}=\langle|u_{1,k}|^2|u_{2,k}|^2\rangle,\quad \sigma_{12} =
\langle|u_{1,k}|^2u_{1,k}u^*_{2,k}\rangle,\quad \sigma_{21} =
\langle|u_{2,k}|^2u_{2,k}u^*_{1,k}\rangle.
\]
The terms with $\sigma_{nm}$ can be neglected, since at the
Brillouin zone edge these coefficients are exactly zero due the
properties of the Bloch waves. Numerics shows that they grow
linearly in $k-k_B$.

The nonlinear generalization of the system
(\ref{EQ10})-(\ref{EQ11}) is derived by using the nonlinear Bloch
waves $\hat{\Psi}_{n,k}\equiv e^{ikx}\hat{u}_{n,k}= \Psi_{n,k}
+\mathcal{O}(c)$ in the expansion (\ref{EQ5}).  For a small
acceleration the term proportional to $\alpha c$ can be neglected,
thus the inter-band coupling coefficient $\kappa_{12}$ in the
nonlinear system is given by the same formula as in the linear
case. Setting now $A_1 = \exp\left\{-i\int^t_0\md
\tau\,F(q(\tau))\right\} a_1$ and $A_2 = \exp\left\{-i\int^t_0\md
\tau\,F(q(\tau))\right\}(-ia_2)$, with $F = (E_1+E_2)/2
+c\chi_{12}$, we obtain the nonlinear two-band model:
\eqb
i\frac{\md a_1}{\md t} = \left\{-\vare(q) + c\gamma_{11}(q)|a_1|^2
 \right\}a_1 + \alpha\kappa_{12}(q)a_2,
\label{EQ14}\eqe
\eqb
i\frac{\md a_2}{\md t} = \left\{\vare(q) +
c\gamma_{22}(q)|a_2|^2\right\}a_2 +\alpha\kappa_{12}(q)a_1.
\label{EQ15}\eqe
Here we have used the modified Bloch energies from equation
(\ref{EQ13}) and defined
\eqb
\gamma_{nn} = \chi_{nn} - \chi_{12} = \langle|u_{n,k}|^4 -
|u_{1,k}|^2|u_{2,k}|^2\rangle ,\quad n=1,2.
\label{EQ16}\eqe

The system (\ref{EQ14})-(\ref{EQ15}) depends on the fine
$k$-structure of the Bloch bands, since $q(t)$ scans through the
edge of the Brillouin zone. Hence, the expressions for the Bloch
waves in a vicinity of the Brillouin zone edge have to be
available. For a general optical lattice   the coefficients
$\vare(q)$, $\kappa_{12}(q)$, and $\gamma_{nn}(q)$ can be computed
numerically. However, such expressions are available in the
analytical form (see below) for a weak optical lattice used in the
experiments \cite{LZ3,LZ4}.

\section{Weak lattice approximation}
\label{SECIII}
Similar to the free electrons approach  in the solid state theory
\cite{Ziman},  in the vicinity of the Brillouin zone edge the
Bloch waves from the two lowest bands of a weak optical lattice
can be sought in the  form
\eqb
\Psi = a e^{iqx} + b e^{i(q-2k_B)x}.
\label{EQ17}\eqe
For the cosine lattice $v(x) = v_0\cos(x)$ we get
\eqb E_{1,2}(q) = \frac{q^2}{2} + k_B(k_B - q) \mp \vare(q),\quad \vare(q)\equiv
\left(k_B^2(k_B-q)^2 +\frac{v_0^2}{4} \right)^{\frac{1}{2}}.
\label{EQ18}\eqe
The corresponding Bloch waves read
\eqb
\Psi_{1,q} =  \frac{\nu(q) e^{iqx} - e^{i(q-2k_B)x}}
{\sqrt{1+\nu^2(q)}}, \quad \Psi_{2,q} = \frac{ e^{iqx} +\nu(q)
e^{i(q-2k_B)x}} {\sqrt{1+\nu^2(q)}},
\label{EQ19}\eqe
where $\quad \nu(q) = 2[k_B(k_B-q) + \vare(q)]/v_0$. Then the
remaining two coefficients of the two-band model are
\eqb
\kappa_{12}(q) = \frac{1}{2\vare(q)}\frac{\nu(q)}{1+ \nu^2(q)},
\quad \gamma(q) \equiv\gamma_{nn}(q)= \frac{4\nu^2(q)}{[1 +
\nu^2(q)]^2}.
\label{EQ20}\eqe

\begin{figure}[ht]
\begin{center}
\includegraphics{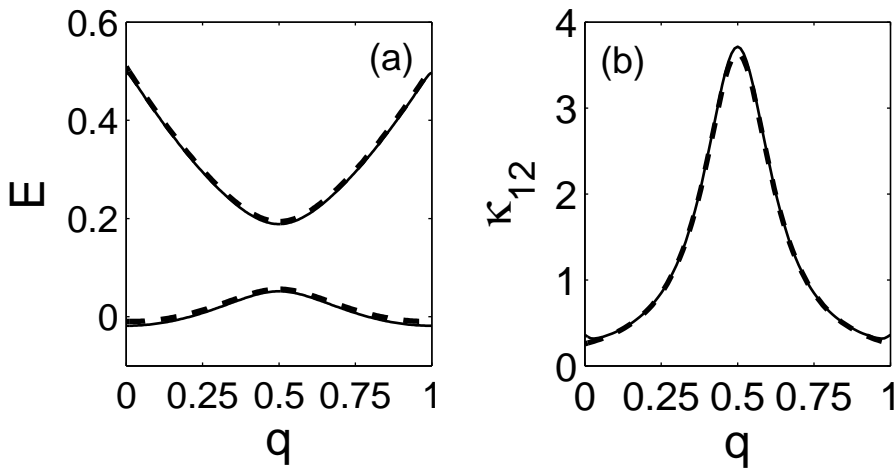}
\caption{\label{FG1} The Bloch energy  of the first two bands,
panel (a), and the inter-band coupling coefficient $\kappa_{12}$,
panel (b), vs. the band index $q$. Numerically found values are
given by the solid lines and the analytical weak-lattice
approximations are given by the dashed lines. Here $v_0=0.1375$.}
\end{center}
\end{figure}

For the linear case, the analytical formulae in equations
(\ref{EQ18}) and (\ref{EQ20}) are in good agreement with the exact
numerical results for the experimentally used values
\cite{LZ3,LZ4} (we use $v_0=0.1375$). In fig. \ref{FG1} the
analytical energy levels $E_{1,2}(q)$ (left panel) and the
inter-band coupling coefficient $\kappa_{12}(q)$ (right panel) vs.
their numerical values are given. The coupling coefficients to the
higher Bloch bands are small  as compared to $\kappa_{12}$, in
accordance with the theory of section \ref{SECII1}. For instance,
we have for the maximum value: $\kappa_{n3}(k_B)<0.2$, $n=1,2$,
for $v_0=0.1375$.

\begin{figure}[ht]
\begin{center}
\includegraphics{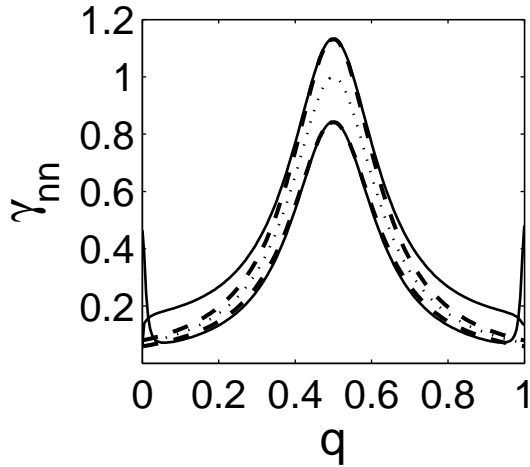}
\caption{\label{FG2} The nonlinear coefficients $\gamma_{11}(q)$
and $\gamma_{22}(q)$. The numerically computed coefficients  are
given by the solid lines and the analytical weak-lattice
approximation is given by the dotted line. The dashed lines are
the scaled analytical coefficients as indicated in the text. Here
$v_0=0.1375$.}
\end{center}
\end{figure}

The weak lattice approximation cannot, however, capture the weak
asymmetry in the two nonlinear coefficients $\gamma_{11}(q)$ and
$\gamma_{22}(q)$, since their analytical values always coincide.
In fact, in the vicinity of the Brillouin zone edge, the
analytical approximation (dotted line in fig. \ref{FG2}) is close
to the algebraic average of the exact numerical coefficients. To
have a better approximation the analytical coefficients must be
scaled by the maximum value (at $q=k_B$). For instance,
$\gamma_{11}(q) = 1.132\gamma(q)$ and $\gamma_{22}(q) =
0.842\gamma(q)$ for $v_0=0.1375$ (the dashed lines in fig.
\ref{FG2}).

\section{Comparison with the Landau-Zener model}
\label{SECIV}

To make a quantitative comparison with the nonlinear Landau-Zener
model
\eqb
i\frac{\md a}{\md t} = \left\{\frac{\alpha t}{2} +
\frac{c}{2}(|b|^2 - |a|^2)\right\}a +\frac{v_0}{2}b,
\label{EQ21}\eqe
\eqb
i\frac{\md b}{\md t} = -\left\{\frac{\alpha t}{2} +
\frac{c}{2}(|b|^2 - |a|^2)\right\}b +\frac{v_0}{2}a,
\label{EQ22}\eqe
introduced in Refs. \cite{WN,ZG}, one has to relate the amplitudes
$a$ and $b$ of the incident and Bragg-scattered  waves  with the
amplitudes of the  Bloch waves from the first two bands. There are
two  approaches. The first one is based on the fact that the
adiabatic energy levels of the quantum-mechanical Hamiltonian
\eqb
H = \left(\begin{array}{cc} \frac{\alpha t}{2} & \frac{v_0}{2}\\
\frac{v_0}{2} & -\frac{\alpha t}{2}\end{array}\right)
\label{EQ23}\eqe
associated with the system (\ref{EQ21})-(\ref{EQ22}), mimic the
structure of the first two Bloch bands. Therefore,  the Bloch
amplitudes are obtained by projecting the vector $(a,b)$ onto the
instantaneous adiabatic eigenvectors of  $H$ \cite{Biao}. We
obtain the following relation
\eqb
a_1 = \frac{\varkappa a - b}{\sqrt{1 + \varkappa^2}},\quad a_2 =
\frac{a +\varkappa b}{\sqrt{1 + \varkappa^2}},
\label{EQ24}\eqe
where  $\varkappa = (E-\alpha t)/v_0$ with $E = \sqrt{\alpha^2 t^2
+v_0^2}$. The amplitudes $a_1$ and $a_2$ in equation (\ref{EQ24})
correspond to the instantaneous adiabatic energy levels $E_{1,2} =
\mp E/2$, thus they can be taken as the occupation amplitudes of
the first two Bloch bands according to the nonlinear Landau-Zener
model (however, one can get only the absolute values of the
amplitudes $a_{1,2}$, their  phases are left unknown due to
arbitrary phases of the eigenvectors).

The second approach is to use the analytical results for a weak
lattice and equate directly the order parameters corresponding to
the two models.    The two-band model (\ref{EQ15})-(\ref{EQ16})
corresponds to the order parameter
\eqb
\psi = e^{i\theta(t)+i\mathrm{v}(t)x}\left[a_1(t)\Psi_{1,q(t)}(x)
-ia_2(t)\Psi_{2,q(t)}(x)\right],
\label{EQ25}\eqe
with $\theta(t)$ being an unimportant phase and $q(t) = q_0 -
\mathrm{v}(t)$. The nonlinear Landau-Zener model
(\ref{EQ21})-(\ref{EQ22}) corresponds to the following order
parameter \cite{WN}
\eqb
\psi = e^{-3ict/2}\left[a(t)e^{ikx} + b(t)e^{i(k-2k_B)x}\right].
\label{EQ26}\eqe
It is straightforward to relate the  band amplitudes using the
expression for the Bloch waves (\ref{EQ19}) and associating
$k=q_0$. The common $x$-independent phase can be discarded. We
have
\eqb
\left(\begin{array}{c}a \\ b\end{array}\right) =
\left(\begin{array}{cc}\frac{\nu}{\sqrt{1+\nu^2}} &
\frac{-i}{\sqrt{1+\nu^2}} \\ \frac{-1}{\sqrt{1+\nu^2}} &
\frac{-i\nu}{\sqrt{1+\nu^2}}
\end{array}\right)\left(\begin{array}{c}a_1\\
a_2\end{array}\right).
\label{EQ27}\eqe
The transformation given by equation (\ref{EQ27}) is linear and
unitary. Asymptotically it trivializes. Indeed,  $\nu\to \infty$
as $t\to -\infty$, while $\nu\to 0$ as $t\to \infty$ (here
$\alpha>0$). We get
\eqb
a(-\infty) = a_1(-\infty),\quad b(-\infty) = -ia_2(-\infty),\quad
a(\infty) = -ia_2(\infty),\quad b(\infty) = -a_1(\infty).
\label{EQ28}\eqe

\begin{figure}[ht]
\begin{center}
\includegraphics{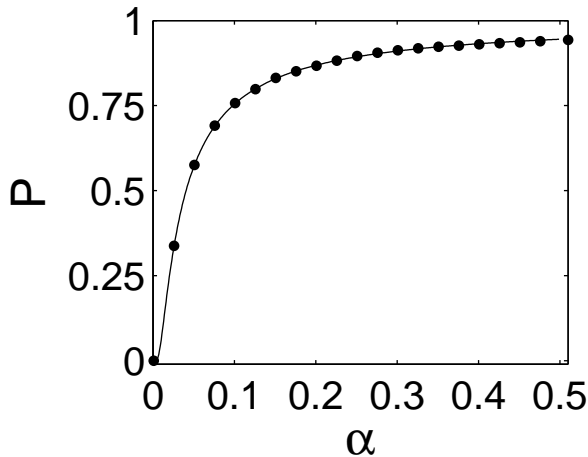}
\caption{\label{FG3} The tunneling probability  vs. the lattice
acceleration according to the Landau-Zener formula (solid line)
and in the two-band system (points) in the case of negligible
nonlinearity and trivial initial condition (i.e. just one band
being populated initially). Here $v_0 = 0.134$ and $c=0$.}
\end{center}
\end{figure}

The overall conclusion to be argued below is that, whatever the way one relates the
two models, the predictions of the nonlinear Landau-Zener model
(\ref{EQ21})-(\ref{EQ22}) and the two-band model (\ref{EQ15})-(\ref{EQ16}) disagree
significantly in general, if the initial condition is non-trivial, i.e. both
amplitudes ($a$ and $b$ and, respectively, $a_1$ and $a_2$) are non-zero initially.
The disagreement is apparent even for a negligible nonlinearity, which case is
treated below in detail. A remarkable agreement between the two models if just one
band is populated initially (i.e. either $a_1$ or $a_2$ is zero initially) must
also be stressed.

\begin{figure}[ht]
\begin{center}
\includegraphics{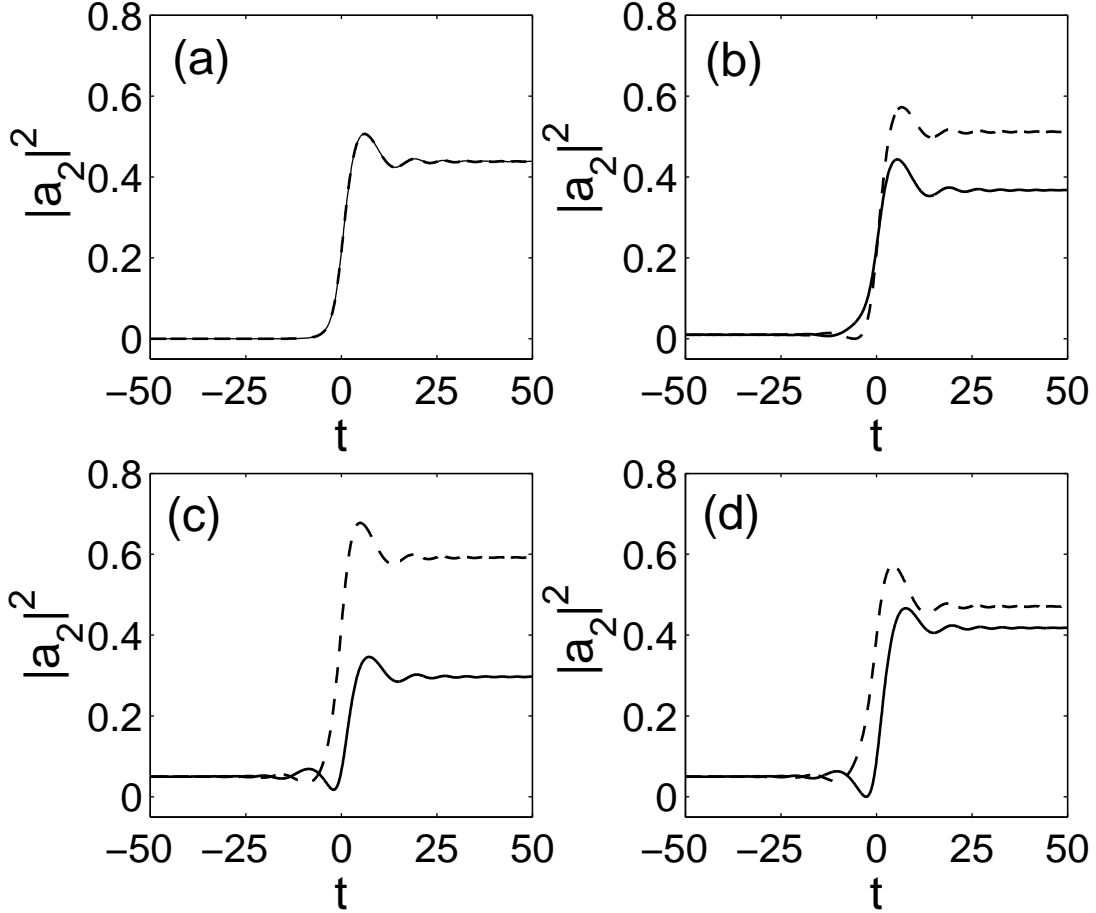}
\caption{\label{FG4} The population of the second Bloch band vs.
time in the two-band model (solid lines) and in the nonlinear
Landau-Zener model (dashed lines). We have used $v_0=0.1375$,
$\alpha =0.036$, and $c=0$ and the following initial conditions:
(a) $(a_1,a_2)= (1,0)$, (b) $(a_1,a_2) =
(\sqrt{0.99},\sqrt{0.01})$, (c) $(a_1,a_2) =
(\sqrt{0.95},\sqrt{0.05})$, and (d) $(a_1,a_2) =
(\sqrt{0.95},e^{-i\pi/5}\sqrt{0.05})$. The initial amplitudes $a$
and $b$ for the  Landau-Zener model and the corresponding
functions $a_1(t)$ and $a_2(t)$ were computed using equation
(\ref{EQ27}). The initial time was $t_0 = -200$.  }
\end{center}
\end{figure}

When initially the atoms populated just one band the two models
give the same results. For instance, for $c=0$ the probability of
tunneling, defined as $P=|a_n(\infty)|^2$ for $a_n(-\infty)=0$, is
given by the Landau-Zener formula $P = e^{-\pi v_0^2/(2\alpha)}$
\cite{LZMOD} which is illustrated in fig. \ref{FG3}.

 When both bands are populated initially, the two models
do not agree in the prediction of the final populations. This is
illustrated in fig. \ref{FG4}, where we plot the time-dependence
of  the the second band population for different initial
conditions but with the majority of atoms populating initially the
first band.  Note that in panel (a) the two models give the same
results if related by equation (\ref{EQ27}). However, the same
equation leads to significant disagreement in panels (b) and (c).
Similar conclusion can be drawn using the first relation scheme
for the two models, i.e. equation (\ref{EQ24}).

Two more conclusions are apparent from fig. \ref{FG4}. When the
first band is initially dominantly populated,  a weak population
of the second band makes a strong effect on its final population,
(compare panels (a) and (c)). Moreover, at the fixed initial
populations, the initial phase difference can have a significant
effect on the final populations (compare panels (c) and (d)).  The
same conclusions are true for the first band.

\begin{figure}[ht]
\begin{center}
\includegraphics{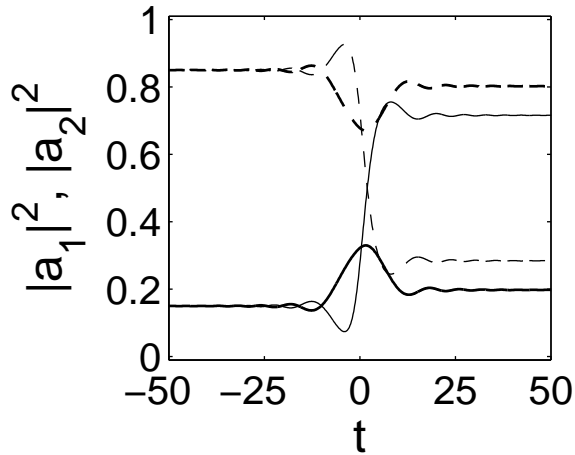}
\caption{\label{FG5} The populations of the  Bloch bands vs. time
in the two-band model (thick lines) and in the Landau-Zener model
(thin lines).  The solid lines give $|a_1|^2$. The initial time
was $t_0 = -200$. Here $v_0=0.1375$, $\alpha = 0.036$, and  $c=0$.
We have used the following initial condition $(a_1,a_2) =
(i\sqrt{0.15}, \sqrt{0.85})$. The initial amplitudes $a$ and $b$
for the  Landau-Zener model and the corresponding functions
$a_1(t)$ and $a_2(t)$ were computed using equation (\ref{EQ27}). }
\end{center}
\end{figure}

The significant disagreement between the two models can be
manifested by using a larger initial population of the less
populated Bloch band. In this case the disagreement is
qualitative, see fig. \ref{FG5}.

The transformation (\ref{EQ27}) can be used to derive the two-band
model in the diabatic basis, i.e. in terms of the amplitudes of
the  incident ($a$) and Bragg scattered ($b$) waves. The resulting
system is given in appendix \ref{APPB}. The main feature of the
two-band model in the diabatic basis is that both the sweep
function and the coupling coefficient are complicated functions of
the band index $q$ with the coupling coefficient being complex
valued. This latter fact can explain the disagreement in the
predictions of the two-band model and the nonlinear Landau-Zener
model when  the two bands are initially populated.


\section{Asymmetric tunneling for $c\ne 0$}
\label{SECV}

\begin{figure}[ht]
\begin{center}
\includegraphics{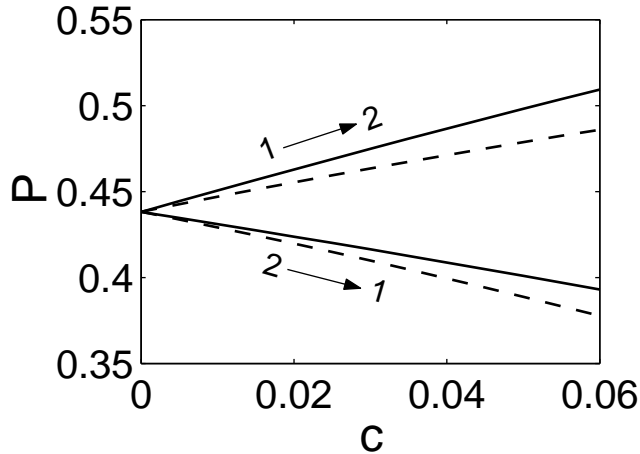}
\caption{\label{FG6} The  tunneling probabilities in the two-band
model (solid lines)  when just one band is initially populated.
The upper line gives the lower-to-upper tunneling probability and
the lower one gives the upper-to-lower tunneling probability. For
comparison, the tunneling probabilities in  the nonlinear
Landau-Zener model are also given (dashed lines). Here
$v_0=0.1375$ and $\alpha = 0.036$.  }
\end{center}
\end{figure}

The tunneling probability  of BEC at the edge of the Brillouin zone in an
accelerated optical lattice was found to have an asymmetric dependence on the
nonlinearity coefficient $c$ \cite{LZ3}. The asymmetry in the tunneling probability
was first theoretically predicted in Ref. \cite{ZG} within the nonlinear
Landau-Zener model. The two-band model (\ref{EQ15})-(\ref{EQ16}) confirms the
existence of the asymmetric tunneling, see fig. \ref{FG6}.

The experimental results (see fig. 2 in Ref. \cite{LZ3}) also suggest that the
upper-to-lower tunneling probability, in contrast to the  lower-to-upper one, is
unaffected by the nonlinearity (there is, however, a large error due to
imperfections in the experiment). Below we argue that, indeed,  there is  an
additional asymmetry in the sensitivity of the tunneling probability to the initial
nonlinearity-induced population of the adjacent Bloch band, which, together with
the asymmetric dependence on the nonlinearity coefficient, can give a complete
explanation of  the experimental results.

In section \ref{SECIV} (see fig. \ref{FG4}) we have found that the final
populations of the Bloch bands are affected by the initial band populations and
phases. Due to imperfections in the experimental preparation of the initial state,
there is always a small population of BEC atoms in the adjacent Bloch band. The
condensate is initially prepared at the center of the Brillouin zone ($k=0$) with a
large fraction of atoms populating one of the two Bloch bands.  Besides  the
experimental imperfections,  nonlinearity (i.e. the atomic interactions in BEC)
also contributes to a small number of BEC atoms in the adjacent band, since the
nonlinearity couples the Bloch bands in the higher orders of the perturbation
theory discarded in section \ref{SECII2}.

\begin{figure}[ht]
\begin{center}
\includegraphics{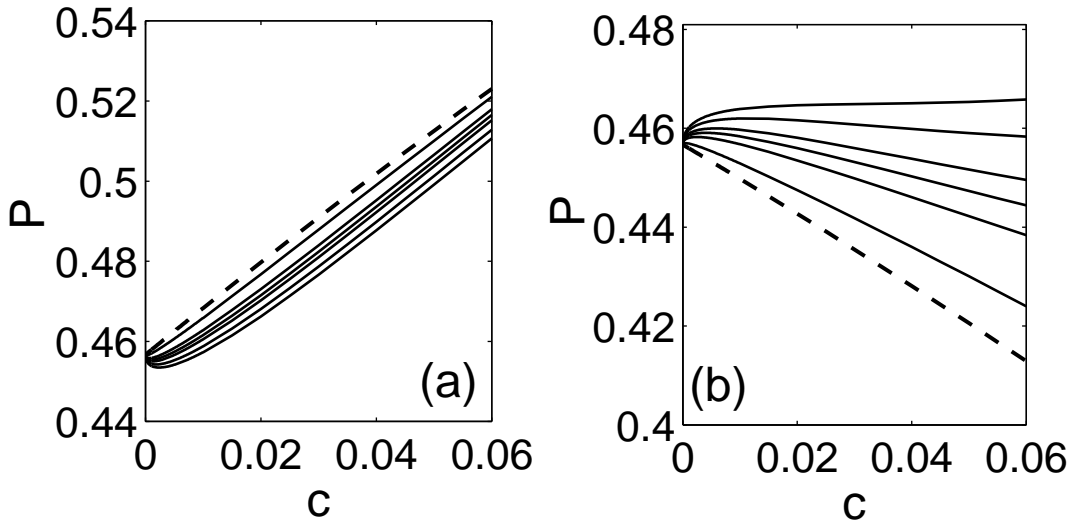}
\caption{\label{FG7} The lower-to-upper (a) and upper-to-lower tunneling  (b) in
the two-band model with a weak nonlinearity-induced initial population of the
adjacent Bloch band (the upper band in case of (a) and the lower one in case of
(b)). The probability of tunneling  (the final population of the Bloch band to
which the tunneling takes place) is given vs. the nonlinearity coefficient $c$. The
initial conditions are (a) $[a_1,a_2] = [\sqrt{1-rc},\sqrt{rc}]$ and (b) $[a_1,a_2]
= [\sqrt{rc},\sqrt{1-rc}]$.  The dashed lines give the tunneling probablity for
$r=0$. The solid lines have $r=\{0.01; 0.05; 0.075; 0.1; 0.15; 0.2\}$ (the solid
lines deviate from the dashed one as $r$ increases). Here $v_0=0.134$ and $\alpha =
0.036$. The time interval is $[-60, 60]$. }
\end{center}
\end{figure}

To find out the effect of the nonlinearity-induced population of the Bloch band on
the tunneling probability to this band, we choose to use the following initial
condition $a_j = \sqrt{rc}$, where $j=2$ for the lower-to-upper tunneling and $j=1$
for the upper-to-lower tunneling (we do not know  the experimental initial share
$|a_j|^2$ of the number of atoms in the adjacent Bloch band, however, this number
grows with the nonlinearity coefficient $c$ \footnote{The particular choice of the
dependence of the initial population in the adjacent band on the nonlinearity
coefficient $c$ is not crucial as long as the population grows with $c$. The point
is that the asymmetry in the dependence of the tunneling probability on the initial
band populations would be qualitatively the same. On the other hand, since we use
the analytical formulae of the weak lattice approximation valid only in the
vicinity of the Brillouin zone edge, we can claim only the qualitative
correspondence with the full GPE equation. }). Since the Bloch waves are real for
$k=0$ the initial phase difference between the amplitudes $a_1$ and $a_2$ of the
two-band model is close to zero and does not affect the general picture. For the
coefficient $r$ we have used the following set of values $\{0.01; 0.05; 0.075; 0.1;
0.15; 0.2\}$. The results of the numerical simulations of the two-band model are
presented in fig. \ref{FG7}. The probability of tunneling (defined here as the
final population of the Bloch band to which the tunneling takes place, neglecting a
very small initial population of this band) is given vs. the nonlinearity
coefficient $c$. The asymmetry in the spread of the numerical curves with different
$r$ in the two cases of tunneling is apparent. Here we should mention that  fig.
\ref{FG7} can  provide only the qualitative account of the asymmetry in the
sensitivity of tunneling to the initial Bloch band populations, since the weak
lattice approximation developed in section \ref{SECIII} applies only in the
vicinity of the Brillouin zone edge.

It is easy to explain the tunneling probability change if the Bloch  band to which
the tunneling takes place  is populated with a small number of atoms. Indeed, the
nonlinear terms in the two-band system (\ref{EQ14})-(\ref{EQ15}) effectively raise
the band energy levels. Therefore, in the case of the lower-to-upper tunneling a
small population of the adjacent band increases the energy gap, while in the case
of the upper-to-lower tunneling it decreases the energy gap, thus the tunneling
probability is decreased in the first case and increased in the second one.
However, this argument cannot explain the fact that the strength of the sensitivity
to the population is quite different in the two cases.

In conclusion, a very small  initial population of the Bloch band to which the
tunneling takes place (in fact for $r=0.2$ we get $|a_j|^2= 0.12$) has a dramatic
effect on its final population in the case of the upper-to-lower tunneling as
compared to that in the case of the lower-to-upper tunneling (if one consider the
relative change in the tunneling probability when increasing $r$). This fact,
together with the known asymmetry of the tunneling probability, provides a complete
explanation of the experimental results presented in Ref. \cite{LZ3}.

\section{Conclusion}
\label{SECVI}
In the present  work  a finite-dimensional model has been derived for the
condensate tunneling at the edge of the Brillouin zone in an accelerating optical
lattice. The fine Bloch band structure was taken into account. A special  attention
was paid to the case of a weak lattice, which could have seemed  a redundant task
due to the existence of the nonlinear Landau-Zener model proposed for this case
previously \cite{WN,ZG}. However, it turns out that the two-band model derived in
this paper and the Landau-Zener model disagree significantly even in the case of a
negligible nonlinearity if both adjacent bands are populated initially. Only in the
special case of just one Bloch band being initially populated the two models  agree
in the prediction of the final populations. In particular, the Landau-Zener result
\cite{LZMOD} is reproduced by our model.

The two-band model confirms the asymmetry in the tunneling probability dependence
on the nonlinearity,  predicted theoretically in Ref. \cite{ZG} and discovered
experimentally in Ref. \cite{LZ3}. However, besides the qualitative agreement
between the theory and the  experiment, a quantitative disagreement was also
revealed. For instance, the experimental upper-to-lower tunneling seems to be
unaffected by the nonlinearity \cite{LZ3}. We have proposed an explanation for this
fact based on the discovered  strong dependence of the final Bloch band populations
on the initial populations. Indeed,  additionally to the asymmetry in the
dependence of the tunneling probability on the nonlinearity strength, we find also
the asymmetry in the sensitivity of the tunneling probability  to the initial
populations: the upper-to-lower tunneling probability is stronger affected by a
variation of the initial Bloch band populations  than the lower-to-upper tunneling
probability. For instance, about one per cent of BEC atoms populating the lower
Bloch band due to nonlinearity-induced Bloch band coupling significantly modifies
the upper-to-lower tunneling probability, whereas the lower-to-upper tunneling
probability is only weakly affected by the same initial population of the upper
Bloch band.

Though it is in the dependence of the final Bloch band populations on the initial
ones where our model disagrees with the nonlinear Landau-Zener model, nevertheless,
both models show  strong dependence of the final Bloch band populations on the
initial populations. Thus an experimental study and/or numerical simulations of the
full GPE equation with various initial Bloch band populations is in order to
confirm the strong dependence of the final populations on the initial conditions
and compare with the two models.

So far we have discussed the so-called ``Landau-Zener regime'' of BEC tunneling
neglecting the spatial modulation of the Bloch wave amplitudes. Recently  the
modulational instability of a Bloch wave amplitude  was found to result in the
irreversibility of tunneling and the asymmetry in the final band populations
\cite{KKSMOD}. A unified approach, capable to cover the two identified regimes, the
``Landau-Zener regime'' and the ``instability regime''  can also be developed. The
solution of this problem is left for a future publication.

\section{Acknowledgements}

This work was supported by the CNPq-FAPEAL grant of Brazil. We are
grateful to Vladimir Konotop and Biao Wu  for illuminating
discussions of the nonlinear Landau-Zener tunneling.

\appendix
\section{The coupling coefficient between the Bloch bands}
\label{APPA}

The inner product $\langle u_{n,q}|\pd_q u_{m,q}\rangle$ can be
easily evaluated by using the properties of the Bloch functions
$u_{n,q}(x)$. For generality,  consider an arbitrary periodic
potential $v = v(x)$. Define the Hermitian operator $\mathcal{L}_q
\equiv \frac{1}{2}(-i\pd_x + q)^2 + v(x)$. We have then
$\mathcal{L}_q u_{n,q} = E_n(q)u_{n,q}$ and, hence
\eqb
\mathcal{L}_q \pd_q u_{n,q} = E_n(q)\pd_q u_{n,q} +\frac{\md
E_n(q)}{\md q}u_{n,q} + i\pd_x u_{n,q} - qu_{n,q}.
\label{A1}\eqe
Using this result  and orthogonality of the functions $u_{n,q}$ we
get
\[
E_n(q)\langle u_{n,q}|\pd_q u_{m,q}\rangle = \langle
u_{n,q}|\mathcal{L}_q|\pd_q u_{m,q}\rangle = E_m(q)\langle
u_{n,q}|\pd_q u_{m,q}\rangle +i\langle u_{n,q}|\pd_x
u_{m,q}\rangle.
\]
Therefore, for $m\ne n$ we obtain
\eqb
\langle u_{n,q}|\pd_q u_{m,q}\rangle  = -i\frac{\langle
u_{n,q}|\pd_x u_{m,q}\rangle }{E_m(q)-E_n(q)}.
\label{A2}\eqe
Now, let us evaluate the product on the r.h.s. of equation
(\ref{A2}). We have
\[
\mathcal{L}_q \pd_x u_{m,q} = E_m(q)\pd_x u_{m,q} -\frac{\md
v(x)}{\md x}u_{m,q}.
\]
Hence,
\[
E_n(q)\langle u_{n,q}|\pd_x u_{m,q}\rangle = \langle
u_{n,q}|\mathcal{L}_q|\pd_x u_{m,q}\rangle = E_m(q)\langle
u_{n,q}|\pd_x u_{m,q}\rangle - \langle u_{n,q}| \frac{\md
v(x)}{\md x}| u_{m,q}\rangle
\]
with the result
\eqb
\kappa_{nm}\equiv \langle u_{n,q}|\pd_q u_{m,q}\rangle =
-i\frac{\langle u_{n,q}|\pd_x u_{m,q}\rangle }{E_m(q)-E_n(q)} =
-i\frac{\langle u_{n,q}|\frac{\md v(x)}{\md x}| u_{m,q}\rangle
}{(E_m(q)-E_n(q))^2},\quad n\ne m.
\label{A3}\eqe

What  is left is to consider the diagonal inner product $\langle
u_{n,q}|\pd_q u_{n,q}\rangle$. Here we use that the potential is
even function: $v(-x) = v(x)$. Then the functions $u_{n,q}(x)$ can
be selected to satisfy  $u^*_{n,q}(x) = u_{n,q}(-x)$. We get
\[
\langle u_{n,q}|\pd_q u_{n,q}\rangle =
\frac{1}{2\pi}\int\limits_{-\pi}^{\pi}\md x\, u_{n,q}(-x)\pd_q
u_{n,q}(x) = \frac{1}{2\pi}\int\limits_{-\pi}^{\pi}\md x\,
\left(-\pd_q  u_{n,q}(-x)\right)u_{n,q}(x)
\]
\[
= \frac{1}{2\pi}\int\limits_{-\pi}^{\pi}\md x\, \left(-\pd_q
u_{n,q}(x)\right)u_{n,q}(-x) = - \langle u_{n,q}|\pd_q
u_{n,q}\rangle,
\]
where to replace the $q$-derivative we have used the independence
of the inner product on $q$: $\langle u_{n,q}|u_{n,q}\rangle = 1$.
Hence $\langle u_{n,q}|\pd_q u_{n,q}\rangle = 0$. A similar
reasoning for $\langle u_{n,q}|(-i)\pd_x u_{m,q}\rangle$ can be
used also to establish that $\kappa_{nm}(q)$ is real and that
$\kappa_{nm}(q) = -\kappa_{mn}(q)$.

\section{The two-band model in the diabatic basis }
\label{APPB}
Unitary transformation (\ref{EQ27}) for the two-band model can be
interpreted as a relation between the adiabatic basis $(a_1,
a_2)$, where the amplitudes are the band populations, and the
diabatic basis $(a,b)$, where the variables have the physical
meaning of the amplitudes of the incident and Bragg scattered
waves. For simplicity, we consider below only the linear case,
$c=0$ (the nonlinearity does not pose any problems, but
complicates the presentation). After simple calculations one
arrives at the following system in the diabatic basis:
\eqb
i\frac{\md a}{\md t} = \mathrm{E}(q)a + \mathrm{K}(q)b,
\label{B1}\eqe
\eqb
i\frac{\md b}{\md t} = -\mathrm{E}(q)b + \mathrm{K}^*(q)a,
\label{B2}\eqe
where the coefficients are as follows
\eqb
\mathrm{E}(q) = \frac{1 - \nu^2(q)}{1 + \nu^2(q)}\vare(q), \quad
\mathrm{K}(q) = \frac{2\nu(q)\vare(q)}{1 +\nu^2(q)} +
\frac{i\alpha}{\vare(q)}\frac{\nu(q)}{1 +\nu^2(q)}.
\label{B3}\eqe
Here the functions $\vare(q)$ and $\nu(q)$ are taken from section
\ref{SECIII}.  (The system (\ref{B1})-(\ref{B3}) was also verified
numerically against the two-band model.)

It is easy to see that $\mathrm{E}(q)$ is an odd function of
$k_B-q$, while $\mathrm{K}(q)$ is an even function of this
variable. The following asymptotic properties of the coefficients
can be easily established. Recalling that $\dot{q}= -\alpha$ we
get as $t\to \pm \infty$:
\eqb
\mathrm{E}(q) \to - \frac{\alpha t}{2}, \quad \mathrm{K}(q) \to
\frac{v_0}{2}.
\label{B4}\eqe
At the Brillouin zone edge  we get $\mathrm{E}(k_B)=0$ and
\mbox{$\mathrm{K}(k_B) = {v_0}/{2} + {i\alpha}/{v_0}$.}

However, an important fact is observed about the system (\ref{B1})-(\ref{B2}): the
coupling coefficient $\mathrm{K}(q)$ is a complex valued function. Therefore, to
obtain an equivalent Landau-Zener model (with a real coupling coefficient), one has
to perform the following time-dependent phase transformation
\eqb
a = {a}^{(\mathrm{LZ})}e^{i\Theta(q)}, \quad b =
{b}^{(\mathrm{LZ})}e^{-i\Theta(q)}, \quad \Theta(q) \equiv
\mathrm{arg}\{\mathrm{K}(q)\}/2.
\label{B5}\eqe
For the phase-transformed amplitudes we obtain:
\eqb
i\frac{\md a^{(\mathrm{LZ})}}{\md t} =
\mathrm{E}^{(\mathrm{LZ})}(q)a^{(\mathrm{LZ})} +
|\mathrm{K}(q)|b^{(\mathrm{LZ})},
\label{B6}\eqe
\eqb
i\frac{\md b^{(\mathrm{LZ})}}{\md t} =
-\mathrm{E}^{(\mathrm{LZ})}(q)b^{(\mathrm{LZ})} +
|\mathrm{K}(q)|a^{(\mathrm{LZ})},
\label{B7}\eqe
where $\mathrm{E}^{(\mathrm{LZ})}(q) =
\mathrm{E}(q)+\alpha\Theta^\prime(q)$ ($\Theta^\prime \equiv \md
\Theta/\md q$).

It is easy to see that $\mathrm{E}^{(\mathrm{LZ})} \to -\alpha t/2\;$ as $\;t\to
\pm\infty\;$ and $\mathrm{E}^{(\mathrm{LZ})}(k_B) = 0$, i.e. we have all the
necessary prerequisites for application of the arguments due to Landau and Zener
\cite{LZMOD} to the system (\ref{B6})-(\ref{B7}). This explains the agreement
between the Landau-Zener model and  the two-band model when just one band is
populated initially.


\begin{thebibliography}{99}
\bibitem{KinTerm} The condition $a_sn_{1D}\ll 1$ means that the transverse kinetic
term is much larger then the nonlinear term, which is a requisite
for an effective factorization of the order parameter. Indeed,
``adding back'' the transverse terms to  equation (\ref{EQ2}) we
arrive at the condition: $\frac{\hbar^2}{2m\ell_\perp^2} \gg
\frac{gn_{1D}}{2\pi\ell^2_\perp}|\psi|^2$. Using $|\psi|^2\sim 1$,
due to the normalization as indicated in the text, and the
expression for $g$ one arrives at $a_sn_{1D}\ll 1$.
\end{thebibliography}
\end{document}